\documentclass{aastex62}

\bibpunct{(}{)}{;}{a}{}{,}

\def\apj{{ApJ}}                 
\def\apjl{{ApJ}}                
\def\apjs{{ApJS}}               
\def\ao{{Appl.~Opt.}}           
             
\def\aap{{A\&A}}

\def\mnras{{MNRAS}}

\def\pasa{{PASA}}

\usepackage{comment}
\usepackage{color,epsfig}

\newcommand{\kms}{$\mbox{km~s}^{-1}$}

\graphicspath{{./}{figures/}}

\received{January 1, 2018}
\revised{January 7, 2018}
\accepted{\today}
\submitjournal{ApJ}

\shorttitle{Six new methanol maser transitions}
\shortauthors{S.\ L. Breen et al.}

\begin{document}

\title{Discovery of six new class II methanol maser transitions, including the unambiguous detection of three torsionally excited lines toward G\,358.931$-$0.030}

\correspondingauthor{Shari Breen}
\email{Shari.Breen@sydney.edu.au}

\author[0000-0002-4047-0002]{S.\ L. Breen}
\affiliation{Sydney Institute for Astronomy (SIfA), School of Physics, University of Sydney, NSW 2006, Australia}

\author[0000-0001-7575-5254]{A.\ M. Sobolev}
\affiliation{Ural Federal University, 19 Mira street, 620002 Ekaterinburg, Russia}

\author[0000-0003-4810-7803]{J.\ F. Kaczmarek}
\affiliation{CSIRO Astronomy and Space Science, Australia Telescope National Facility, Box 76, Epping, NSW 1710, Australia}

\author[0000-0002-1363-5457]{S.\ P. Ellingsen}
\affiliation{School of Natural Sciences, University of Tasmania, Private Bag 37, Hobart, Tasmania 7001, Australia}

\author[0000-0001-9525-7981]{T.\ P. McCarthy}
\affiliation{School of Natural Sciences, University of Tasmania, Private Bag 37, Hobart, Tasmania 7001, Australia}

\author[0000-0002-4931-4612]{M.\ A. Voronkov}
\affiliation{CSIRO Astronomy and Space Science, Australia Telescope National Facility, Box 76, Epping, NSW 1710, Australia}

\begin{abstract}

We present the unambiguous discovery of six new class II methanol maser transitions, three of which are torsionally excited (v$_t$=1). The newly discovered 6.18-GHz 17$_{−2}$ $\rightarrow$ 18$_{−3}$ E (v$_t$=1), 7.68-GHz 12$_4$ $\rightarrow$ 13$_3$ A$^-$ (v$_t$=0), 7.83-GHz 12$_4$ $\rightarrow$ 13$_3$ A$^+$ (v$_t$ = 0), 20.9-GHz 10$_1$ $\rightarrow$ 11$_2$ A$^+$ (v$_t$=1), 44.9-GHz 2$_0$ $\rightarrow$ 3$_1$ E (v$_t$=1) and 45.8-GHz 9$_3$ $\rightarrow$ 10$_2$ E (v$_t$=0) methanol masers were detected towards G\,358.931$-$0.030, where the known 6.68-GHz maser has recently been reported to be undergoing a period flaring. The detection of the v$_t$=1 torsionally excited lines corroborates one of the missing puzzle pieces in class II maser pumping, but the intensity of the detected emission provides an additional challenge, especially in the case of the very highly excited 6.18-GHz line. Together with the newly detected v$_t$ = 0 lines, these observations provide significant new information which can be utilised to improve class II methanol maser modelling. We additionally present detections of 6.68-, 19.9-, 23.1- and 37.7-GHz class II masers, as well as 36.2- and 44.1-GHz class I methanol masers, and provide upper limits for the 38.3- and 38.5-GHz class II lines. Near simultaneous Australia Telescope Compact Array (ATCA) observations confirm that all 10 of the class II methanol maser detections are co-spatial to $\sim$0.2 arcsec, which is within the uncertainty of the observations. We find significant levels of linearly polarised emission in the 6.18-, 6.67-, 7.68-, 7.83-, 20.9-, 37.7-, 44.9- and 45.8-GHz transitions, and low levels of circular polarisation in the 6.68-, 37.7- and 45.8-GHz transitions.

\end{abstract}



\section{Introduction} \label{sec:intro}

More than 30 methanol maser transitions have now been detected in the regions surrounding young high-mass stars. These methanol maser transitions have been empirically divided into two classes: class I and class II \citep[e.g.][]{Batrla-1987,Menten-1991}, reflecting the respective pumping schemes and therefore their relationship to the exciting object. Class I methanol masers are pumped via collisions \citep[e.g.][]{Leurini-2016,Sobolev-2018} and so tend to be associated with the shocked gas surrounding outflows and expanding H{\sc ii} regions \citep[e.g.][]{Kurtz-2004,Voronkov-2014}, whereas class II methanol masers are pumped via infrared radiation \citep[e.g.][]{Sobolev-1997a,Cragg-2005} and are exclusively found co-located with young high-mass stars \citep[e.g.][]{Minier-2003,Xu-2008,Breen-2013}.

Class II methanol masers at 6.68-GHz have been studied extensively, and a large portion of the Galactic plane has been completely searched for this transition in the Methanol Multibeam (MMB) Survey \citep{Green-2009}, resulting in the detection of 972 sources \citep{Breen-2015}. Further, targeted searches have discovered a total of 18 class II methanol maser transitions \citep[][and references therein]{Zinchenko-2017,Ellingsen-2012} some of which are relatively common like the 12.2-GHz transition \citep[e.g. associated with 45 per cent of 6.68-GHz MMB masers;][]{Breen-2016}, while other lines with reasonable observed sample sizes \citep[e.g. 19.9-, 23.1-, 37.7-, 38.3-, 38.5-, 85.5-, 86.6-, 107.0- and 156.6-GHz;][]{Ellingsen-2003,Ellingsen-2004,Cragg-2004,Caswell-2000,Ellingsen-2011} have revealed lower detection rates, reflecting that the slightly different specific conditions required to produce strong maser emission in these transitions \citep[e.g.][]{Cragg-2005} are less prevalent that those required for the more readily-inverted and common 6.68- and 12.2-GHz transitions.

Studies of class I methanol maser emission have been mostly limited to targeted observations in the 36.2-, 44.1-, 84- and 95-GHz transitions \citep[e.g.][]{Kurtz-2004,Ellingsen-2005,Cyganowski-2009,Chen-2011,Voronkov-2014,Breen-2019}, but some complete searches for the 36.2- and 44.1-GHz transitions exist, revealing large numbers of maser sources \citep[e.g.][]{YZ-2013,Jordan-2015,Jordan-2017}.

Current class II methanol maser pumping theories are able to broadly account for the observed methanol maser transitions and properties detected in prominent maser sources \citep[e.g.][]{Cragg-2005}. In order to reproduce the observed brightness of the common and strong 6.68- and 12.2-GHz class II methanol masers, one of the key requirements of the current models involves pumping through the first two torsionally excited levels \citep[e.g.][]{Sobolev-1994,Sobolev-1997a}, and maser candidates have been identified in these levels \citep[e.g. the 20.9-GHz 10$_1$ $\rightarrow$ 11$_2$ A$^+$ (v$_t$=1) and the 44.9-GHz 2$_0$ $\rightarrow$ 3$_1$ E (v$_t$=1) transitions;][]{Sobolev-1997b,Cragg-2005}. Despite searches for these torsionally excited lines \citep[e.g.][]{Voronkov-2002,Menten-1986} definitive evidence of maser amplification has not been found (however both works suggested on the basis of indirect arguments that there might be weak maser emission from these transitions toward W3(OH), but no conclusive evidence was available, and neither instance has ever been confirmed) and so the involvement of torsional levels has remained an open question in the maser pumping theories. In addition to the elusive torsionally excited transitions, many other class II methanol maser lines have been predicted to arise under plausible physical parameter combinations, but are yet to be confirmed by observations \citep[e.g.][]{Sobolev-1997b,Cragg-2005}. 

The 6.68-GHz methanol maser, G\,358.931$-$0.030, was first detected in the MMB survey with a peak flux density of 10~Jy in early 2006 \citep[ATCA follow-up observations of the survey detection were performed on 2006 Mar 31;][]{Caswell-2010}. Parkes 64-m telescope observations of this maser found no accompanying 12.2-GHz emission in either 2008 Jun or Dec with 5-$\sigma$ detection limits of 0.75 and 0.80~Jy, respectively \citep{Breen-2012a}, but ATCA observations detected a 0.7~Jy 22-GHz water maser \citep{Titmarsh-2016}. Monitoring observations of the 6.68-GHz methanol maser emission have been conducted with the Hitachi 32-m telescope since 2013 Jan 2, showing the source to maintain a flux density below 10~Jy, until exhibiting an increase in peak flux density from 2019 Jan 14 \citep{Coconuts-2019}, signifying the start of the maser burst. Previous similar maser flaring events \citep[e.g. in S255IR-NIRS3 and NGC6334;][]{Fugisawa-2015,MacLeod-2018} have been linked to episodic accretion \citep[e.g.][]{Moscadelli-2017,Hunter-2017,Hunter-2018} and therefore provide an excellent opportunity to study the mechanisms of high-mass star formation. Using maser monitoring data from a small number of sources (S255IR$-$NIRS3, G\,24.329+0.144, and Cepheus A), \citet{Rajabi-2019} suggested that superradiance could play an important role in strong 6.68-GHz methanol maser flaring events. The current flaring event in G\,358.931$-$0.030 shows rapid variations of numerous maser transitions across a wide range of frequencies and flux densities, providing a challenge for both maser pumping and superradiance theories. Observations of the variability and presence of different maser transitions during a period of bursting activity can therefore provide stringent tests of current theories of the flares in the lines with very high brightness temperatures.

Here we present the results of targeted observations towards G\,358.931$-$0.030 for methanol transitions that have been predicted to produce class II methanol masers, along with one line that is not present in the list of maser candidate transitions, and discuss the implications of a number of new discoveries. The monitoring observations that first discovered the flaring event \citep{Coconuts-2019} form part of a wider effort by the Maser Monitoring Organisation (M2O), a collaborative international program now monitoring a large number of maser sources and triggering following up observations of flaring events \citep[e.g.][]{Burns-2019}. The current observations are just one component of a large program following up the maser flare, targeting G\,358.931$-$0.030 across a broad range of wavelengths, including using Very Long Baseline Interferometry (VLBI) observations and single-dish monitoring of the maser lines reported here.

\section{Observations and Data reduction} \label{sec:obs}

A series of spectral line observations targeting the 6.68-GHz methanol maser G\,358.931$-$0.030 \citep[J2000 position: 17$^h$43$^m$10.02$^s$ $-$29$^\circ$51$'$45.8$''$;][]{Caswell-2010} were undertaken with the Australia Telescope Compact Array (ATCA) on 2019 March 5, 6 and 7 following successful preliminary Mopra observations conducted on 2019 March 2 and 3 (when emission in the 20.9-, 44.9-, 45.8-GHz lines where first discovered, and the 36.2-, 37.7-, 44.1-GHz lines were first detected towards this source). In this paper we limit our results and discussion to the 12 methanol transitions summarised in Table~\ref{tab:obs}. 

The ATCA was in the H214 array configuration, but was in scheduled maintenance which restricted the time and number of antennas (a maximum of 4) available for observations (see Table~\ref{tab:obs} for a summary). A total of four frequency setups were used over three days, comprising a total of 49 spectral zoom windows in the 19.9 to 25.1~GHz and 41.8 to 45.9~GHz ranges on 2019 March 5, the 5.0 to 8.4~GHz range on 2019 March 6 and the 34.4 to 39.3~GHz range on 2019 March 7. The Compact Array Broadband Backend \citep[CABB;][]{Wilson-2011} was configured in CFB 64M-32k mode on both March 5 and 7, allowing a series of 64~MHz zoom bands, each with 2048 spectral channels, providing adequate velocity resolution (0.20 - 0.47~\kms) and coverage ($>$ 418~\kms) for millimetre observations. On March 6 CABB was configured in CFB 1M-0.5k mode, allowing for higher velocity resolution from 1~MHz zooms, each with 2048 spectral channels. Multiple 1~MHz zooms were concatenated in order to provide sufficient velocity coverage ($>$89~\kms) for the centimetre observations. Each of the frequency set ups allowed for an additional two $\times$ 2~GHz continuum bands which were centred on 20.7, 24.09, 42.75 and 45.0 on March 5, 6.0 and 7.5 on March 6 and 35.3 and 38.4~GHz on March 7.

At all frequencies, observations of G\,358.931$-$0.030 were interspersed with observations of a nearby phase calibrator at least every 10 mins; B1714$-$336 was observed during the 5.0 to 8.4~GHz observations, and B1741$-$312 was observed for all other frequencies. All observations were made as a series of cuts over an hour angle range of at least 5 hours, and at all frequencies above 19.9~GHz, pointing corrections were made once per hour. Observations of PKS\,B1253$-$055 and PKS\,B1934$-$638 were carried out for bandpass and primary flux density calibration. 

Data were reduced using the {\sc miriad} software package \citep{miriad}. For each observing epoch, a flux model was fit to the broadband PKS\,B1253$-$055 continuum data, which was then bootstrapped to PKS\,B1934$-$638 for absolute flux density scaling. The resulting flux density model of PKS\,B1253$-$055 was then used to determine the bandpass and amplitude scaling of individual zooms. On 2019 March 6, B1741$-$312 had sufficient parallactic angle coverage to calculate a robust leakage solution. Polarisation observations on this day are accurate to within $\sim0.1$\%  of Stokes I \citep{atca_polarisation}. However, due to insufficient parallactic angle coverage of our phase calibrator on March 5 and 7, PKS\,B1934$-$638 was used to calculate the leakage terms and was assumed to have zero polarised flux at all observed frequencies. This is not absolutely known, though comparison between leakage solutions from PKS\,B1934$-$638 and B1741$-$312 on 2019 March 6 differ by $0.3\%$, giving credence to the assumed polarised flux density of PKS\,B1934$-$638. Although we are confident in our calibration, we recognise that there small number of antennas, relatively low sensitivity, and imperfect parallactic angle coverage could have introduced small errors and so we have been relatively conservative with our polarisation detection limits, only listing detections when they are $>$0.5\% of Stokes I. The absolute uncertainty in primary flux density calibration is expected to be within 10\%.                                                                                                                 

Given the ATCA was in an H214 array, data from antenna 6 (located more than 4~km from all other antennas) was excluded in the imaging of lines at frequencies above 19~GHz, leaving only three baselines between 82 and 138~m. Observations between 5.0 and 8.4~GHz were imaged using all three available baselines for that epoch (i.e. including antenna 6), ranging between 138 and 4408~m. Image cubes of each line were created and used to extract absolute positions at the methanol maser peak velocity channel. \citet{Caswell-1997} estimated that RMS positional uncertainty of ATCA maser observations conducted in this manner to be $\sim$0.4 arcsec (Table~\ref{tab:detections} shows that the differences in the absolute positions measured for the class~II transitions are generally less than $\sim$0.2 arcsec even at high frequencies). This was followed by an iteration of phase-only self-calibration using the model created from the clean components in the initial cubes. In some cases, further phase-only self-calibration iterations were required, reflecting the incomplete arrays utilised in these observations. Following self-calibration, Stokes I, Q, U and V images were created, and used to make Stokes I as well as circularly and linearly polarized intensity spectra. The velocity resolution of the images (and therefore spectra) are given in Table~\ref{tab:obs} and are the native resolution of the observations in all cases except for the 6.18-, 6.68-, 7.68- and 7.83-GHz transitions, the data for which have been smoothed to a velocity resolution of 0.05~\kms. 

A summary of the methanol transitions, the adopted rest frequencies, epoch of observation, the ATCA antennas available on each day, velocity coverage and resolution (after smoothing in the case of the 6.18-, 6.68-, 7.68- and 7.83-GHz transitions), as well as the resultant synthesized beam sizes and RMS noise levels are given in Table~\ref{tab:obs}. 

\begin{table*}

 \caption{Target spectral lines, followed by the adopted rest frequency with errors in the last digit in parenthesis (including superscript references as follows: 1: \citet{Muller-2004}; 2: \citet{Xu-1997}; 3: \citet{Tsunekawa-1995}; and 4: \citet{Pickett-1998}), observation epoch (YYMMDD), the available ATCA antennas (where (6) indicates that antenna 6 was included in the array but excluded from imaging), observed velocity coverage (approximately centred on $-$16~\kms with respect to the LSR), resultant velocity resolution (after smoothing in the case of the four 6 and 7~GHz transitions), the synthesised beam size and the resultant RMS noise of the observations.}
  \begin{tabular}{llclccccl} \hline
 \multicolumn{1}{l}{\bf Spectral line} &\multicolumn{1}{c}{\bf Rest freq.}  	&	\multicolumn{1}{c}{\bf Epoch} 	&	\multicolumn{1}{c}{\bf ATCA}	 & \multicolumn{1}{c}{\bf V$_{coverage}$} & \multicolumn{1}{c}{\bf V$_{res.}$} &	\multicolumn{1}{c}{\bf Synth. beam}&\multicolumn{1}{c}{\bf RMS}\\
 &\multicolumn{1}{c}{\bf (MHz)}  	&	
 & \multicolumn{1}{c}{\bf antennas}	 & \multicolumn{1}{c}{\bf (\kms)} & \multicolumn{1}{c}{\bf (\kms)} &	\multicolumn{1}{c}{\bf ($''$ $\times$ $''$)}& \multicolumn{1}{c}{\bf noise (mJy)}\\ \hline
CH$_3$OH 17$_{−2}$ $\rightarrow$ 18$_{−3}$ E (v$_t$=1) & 6181.128(21)$^4$ & 190306 & 4,5,6 & 97 & 0.05 & 3.0 $\times$ 0.9 & 38 \\  

CH$_3$OH 5$_1$ $\rightarrow$ 6$_0$ A$^{+}$ (v$_t$=0)      & 6668.5192(8)$^1$  & 190306 & 4,5,6 & 89 & 0.05 & 2.8 $\times$ 0.8 & 34\\
CH$_3$OH 12$_4$ $\rightarrow$ 13$_3$ A$^-$ (v$_t$=0)& 7682.232(50)$^3$ & 190306 & 4,5,6 & 97 & 0.05  & 2.4 $\times$ 0.7 & 31\\ 
CH$_3$OH 12$_4$ $\rightarrow$ 13$_3$ A$^+$ (v$_t$=0)& 7830.864(50)$^3$& 190306 & 4,5,6 & 95  & 0.05 & 2.3 $\times$ 0.7 & 32\\ 
\\

CH$_3$OH 2$_1$ $\rightarrow$ 3$_0$ E (v$_t$=0) & 19967.3961(2)$^1$ & 190305 & 3,4,5,(6)& 961 & 0.47   & 11.4 $\times$ 5.0 & 20\\
CH$_3$OH 10$_1$ $\rightarrow$ 11$_2$ A$^+$ (v$_t$=1) &  20970.651(50)$^3$ & 190305 & 3,4,5,(6) & 914 & 0.45 & 10.7 $\times$ 4.6 & 35\\
CH$_3$OH 9$_2$ $\rightarrow$ 10$_1$ A$^+$ (v$_t$=0) & 23121.0242(5)$^1$ & 190305 & 3,4,5,(6) & 829 & 0.41 & 9.3 $\times$ 4.3 & 39\\
\\
CH$_3$OH 4$_{-1}$ $\rightarrow$ 3$_0$ E  &  36169.290(14)$^2$	& 190307 & 3,4,5,(6)	& 530 & 0.26 & 6.0 $\times$ 2.9 & 20 \\
CH$_3$OH 7$_{-2}$ $\rightarrow$ 8$_{-1}$ E (v$_t$=0)	&  37703.696(13)$^2$	& 190307 & 3,4,5,(6) & 763 & 0.25 & 6.3 $\times$ 2.5 & 19\\
CH$_3$OH 6$_2$ $\rightarrow$ 5$_3$ A$^-$ (v$_t$=0) & 	38293.292(14)$^2$	&  190307 & 3,4,5,(6) & 500 & 0.24 & 6.2 $\times$ 2.5 & 22\\
CH$_3$OH 6$_2$ $\rightarrow$ 5$_3$ A$^+$ (v$_t$=0) & 38452.652(14)$^2$ &190307 & 3,4,5,(6) & 498 & 0.24 & 6.2 $\times$ 2.5 & 22\\
\\

CH$_3$OH 7$_0$ $\rightarrow$ 6$_1$ A$^+$ 	    &	44069.410(10)$^1$	& 	190305	& 3,4,5,(6) &435 & 0.21 & 5.1 $\times$ 2.2 & 47	\\
CH$_3$OH 2$_0$ $\rightarrow$ 3$_1$ E (v$_t$=1) & 44955.807(50)$^3$ & 190305 & 3,4,5,(6) & 426 & 0.21 & 5.0 $\times$ 2.1 & 48\\ 
CH$_3$OH 9$_3$ $\rightarrow$ 10$_2$ E (v$_t$=0) & 45843.519(50)$^3$ & 190305 & 3,4,5,(6) & 418 & 0.20  & 4.9 $\times$ 2.2 & 50\\
\hline
\end{tabular}\label{tab:obs}
\end{table*}

\section{Results} \label{sec:results}
ATCA observations of the 14 methanol transitions (listed in Table~\ref{tab:obs}) towards 6.68-GHz methanol maser G\,358.931$-$0.030 have resulted in the detection of emission in 12 of these transitions, six of which are new class~II methanol maser transitions, including the discovery of the first torsionally excited methanol masers. The six new class~II methanol maser transitions are those at frequencies of 6.18-, 7.68-, 7.83-, 20.9-, 44.9- and 45.8-GHz (indicated by bold font in Table~\ref{tab:detections}).  Here we also report the first detection of 36.2-GHz class I methanol maser emission and 37.7-GHz class II maser emission toward this source. We made independent detections of the 19.9-, 23.1- and 44.1-GHz transitions but subsequently found that these had been discovered slightly earlier than our observations \citep[e.g.][MacLeod et al. in prep, Kim et al. in prep]{Volvach-2019}.


The properties of the detected maser lines are provided in Table~\ref{tab:detections}, including the fitted position at the maser peak velocity, flux density and velocity information, along with percentage of linear and circular polarization. As shown, we find only three examples of circularly polarized emission, the most significant ($\sim$1.5\% of the Stokes I flux density) in the 45.8~GHz transition, but find significant linear polarization ($>$0.5\% of the Stokes I flux density) to be relatively common.

\begin{figure*}
\epsfig{figure=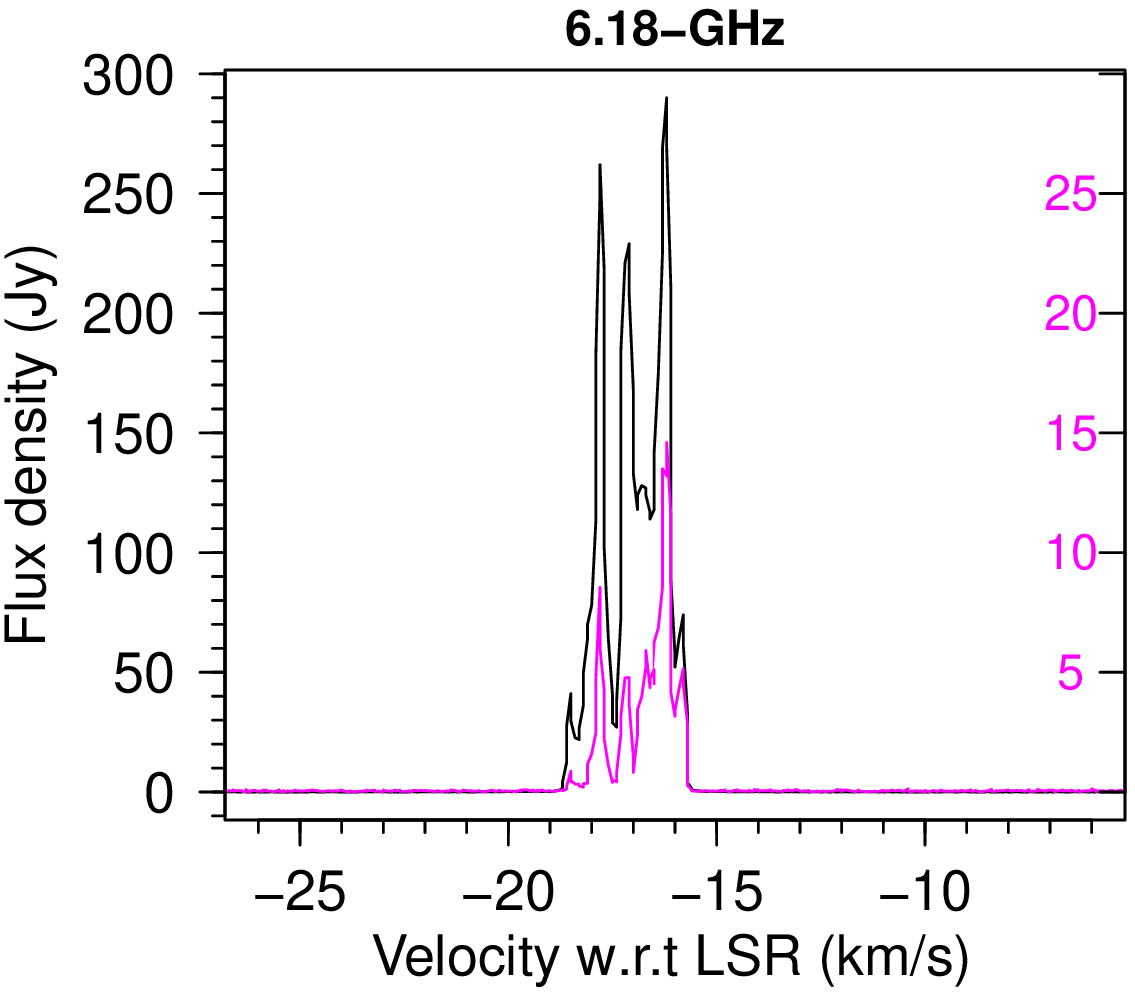,scale=0.51,angle=270}\hspace{-0.4cm}
\epsfig{figure=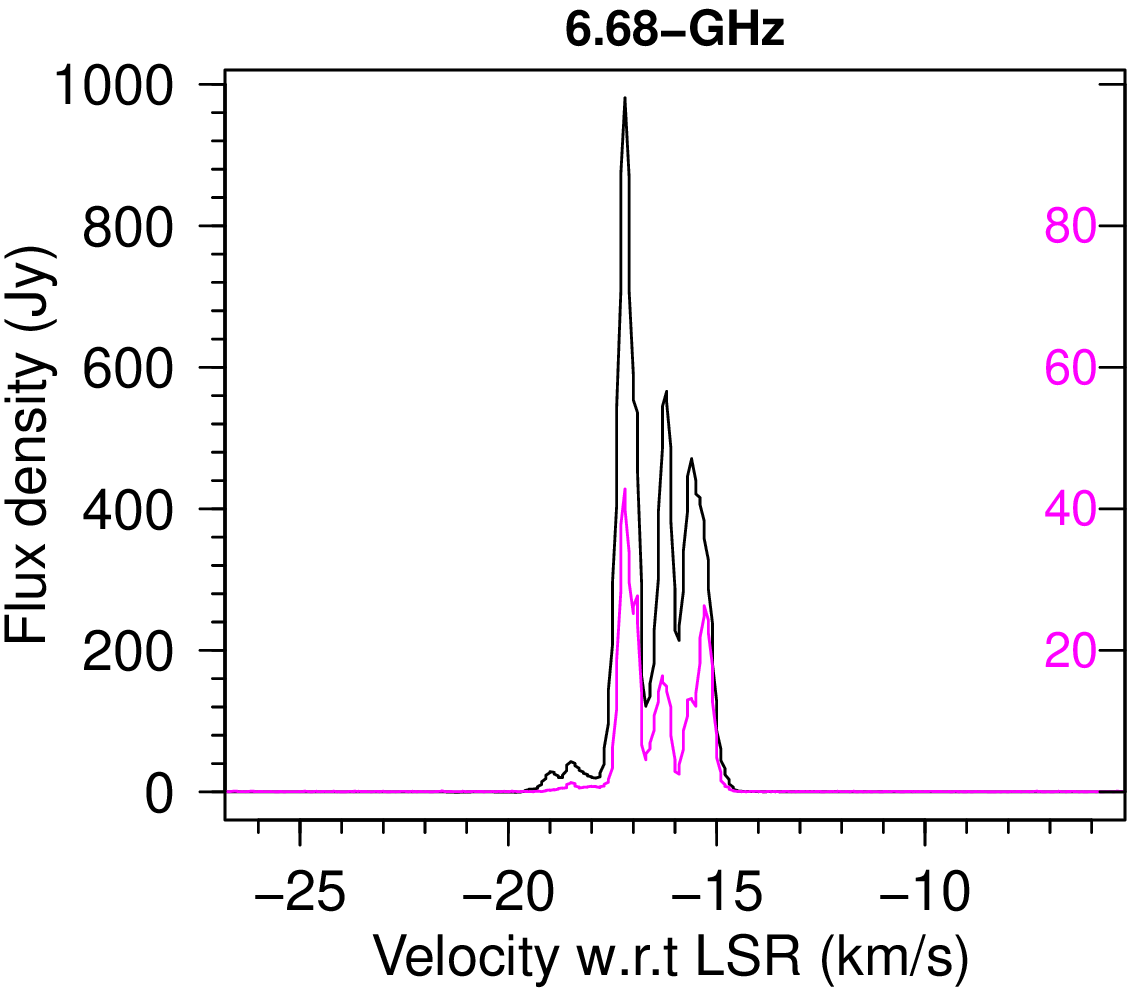,scale=0.51,angle=270}\hspace{-0.4cm}
\epsfig{figure=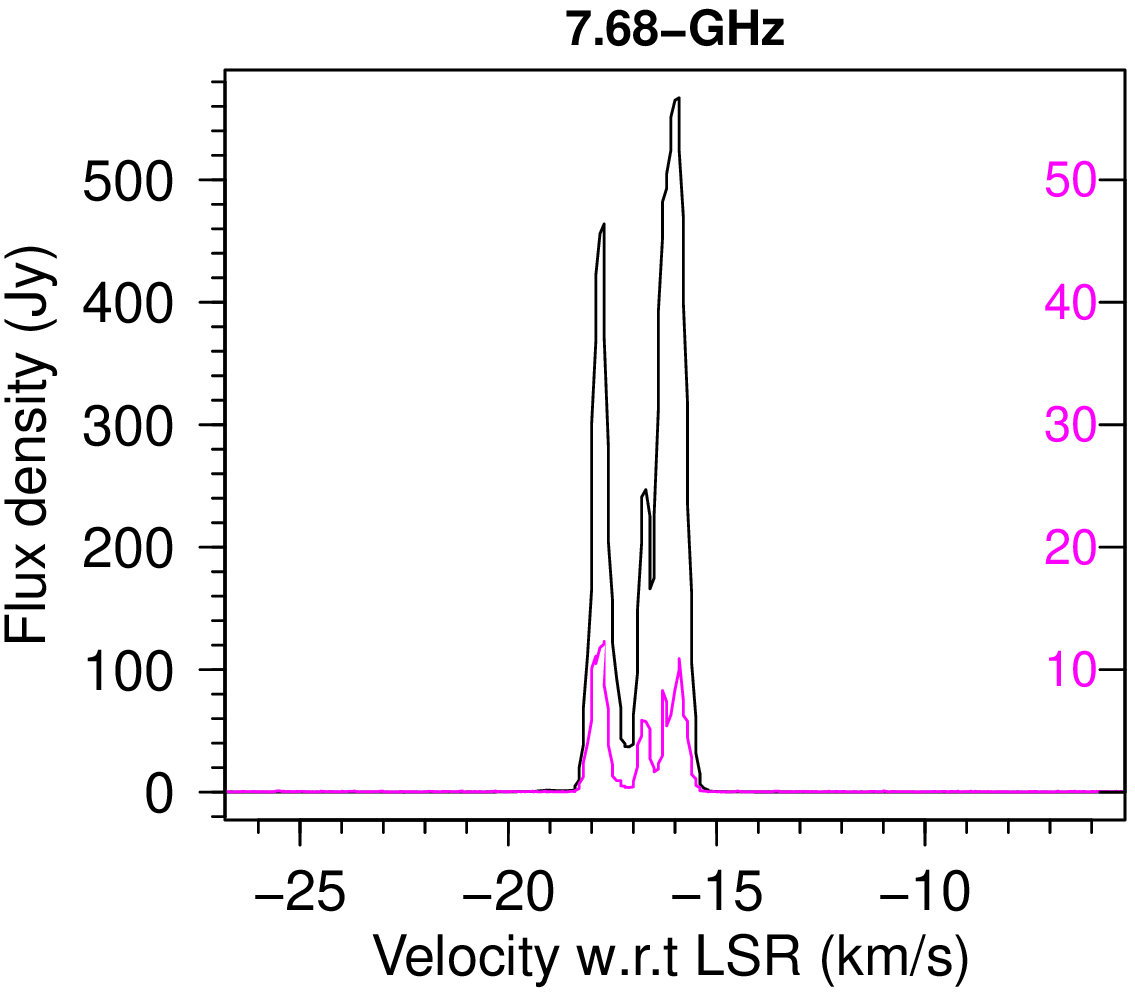,scale=0.51,angle=270}\vspace{-1cm}
\epsfig{figure=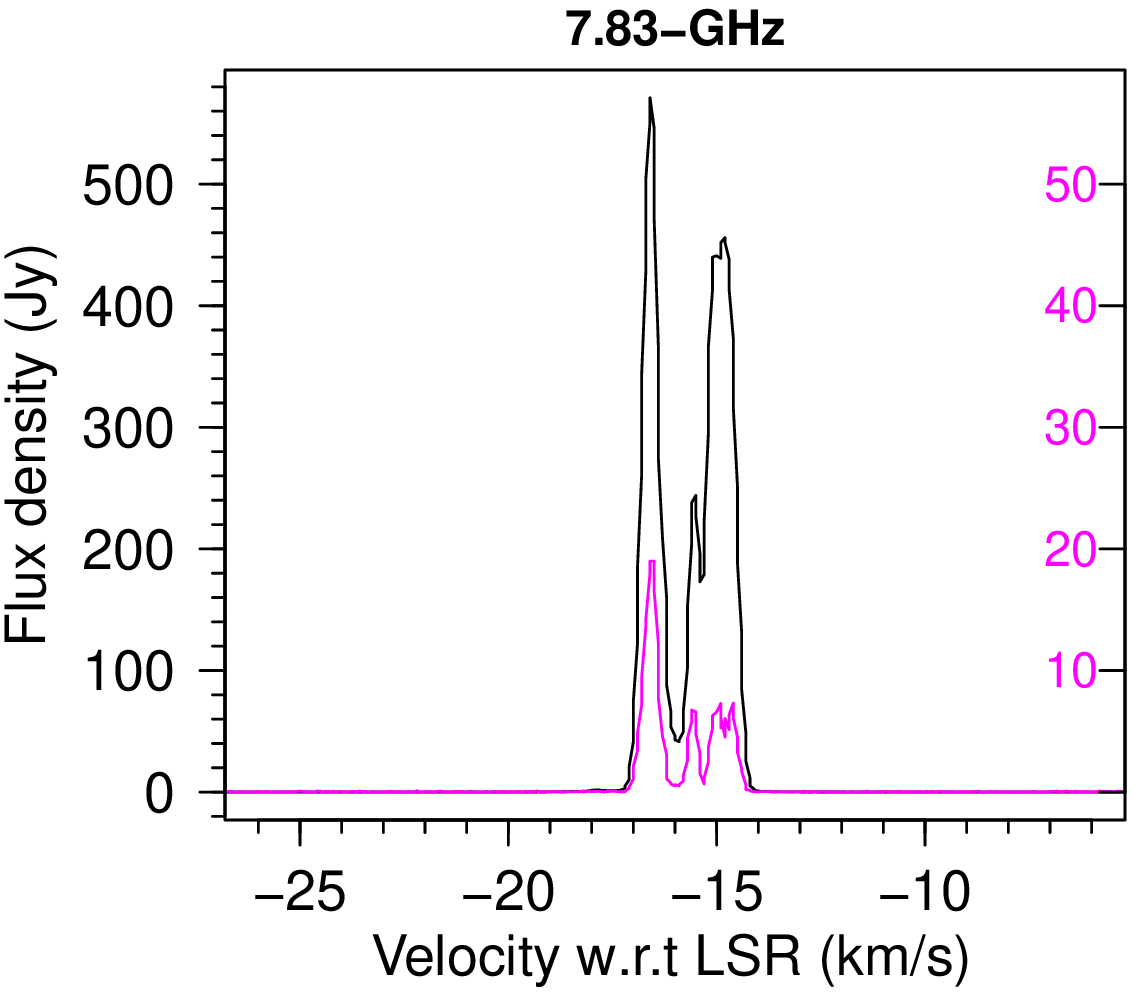,scale=0.51,angle=270}\hspace{-0.4cm}
\epsfig{figure=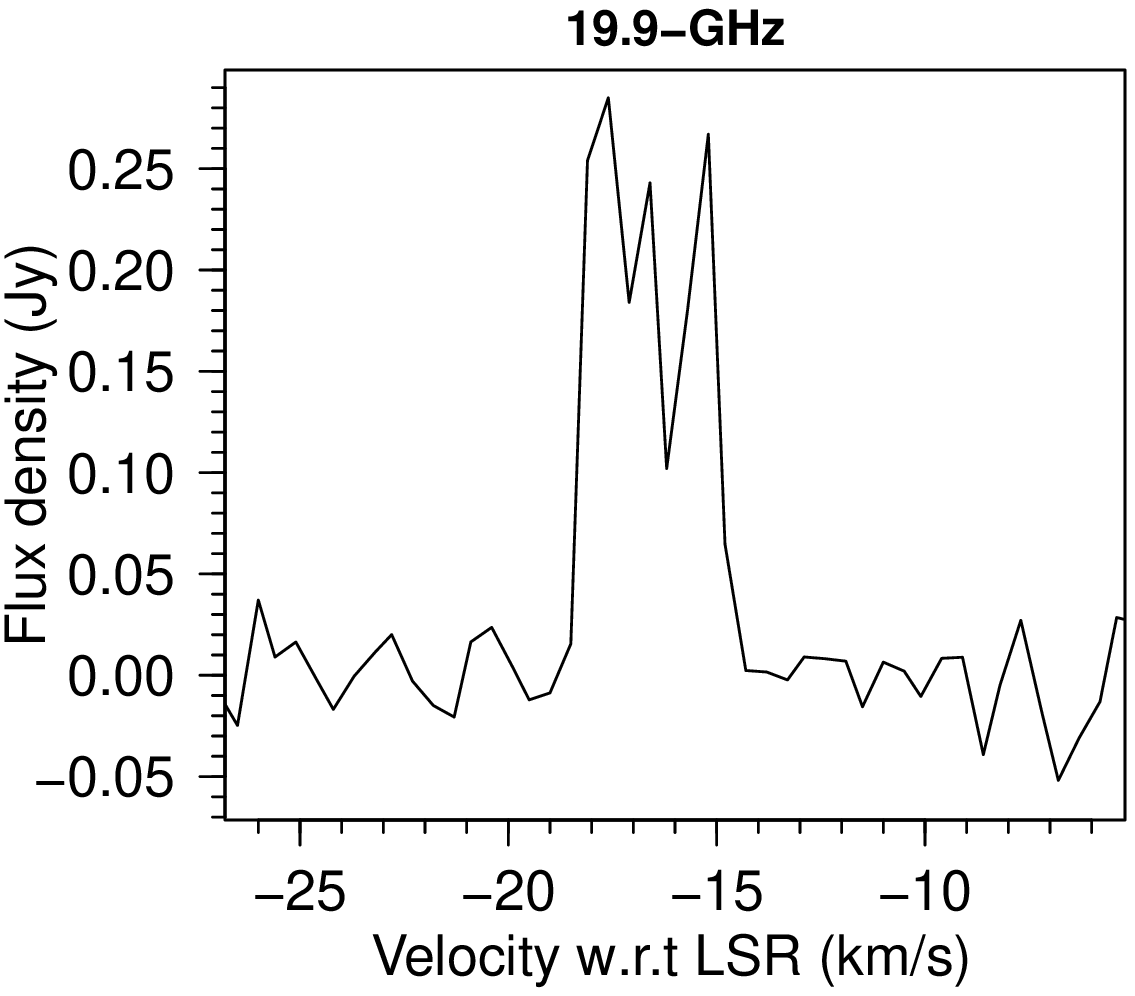,scale=0.51,angle=270}\hspace{-0.4cm}
\epsfig{figure=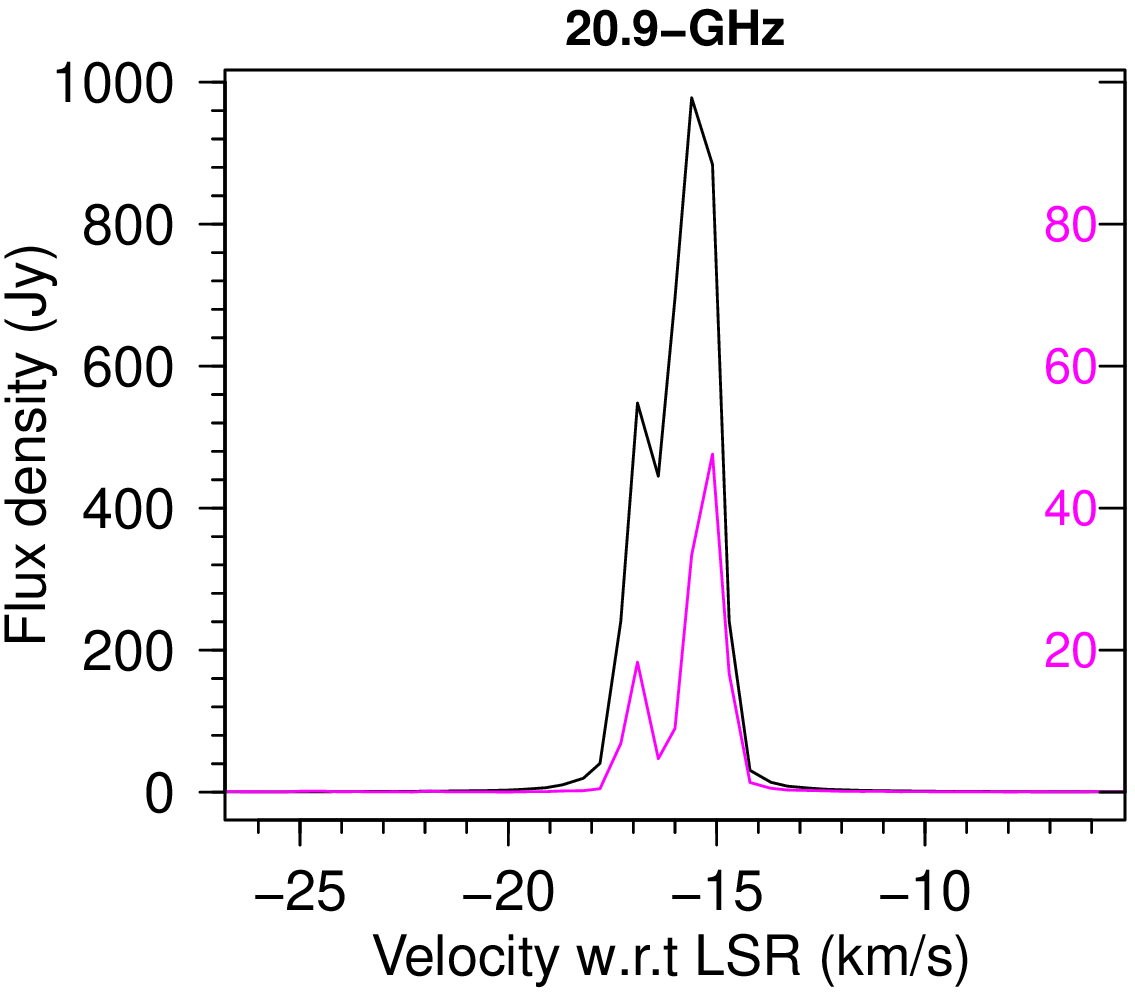,scale=0.51,angle=270}\vspace{-1cm}
\epsfig{figure=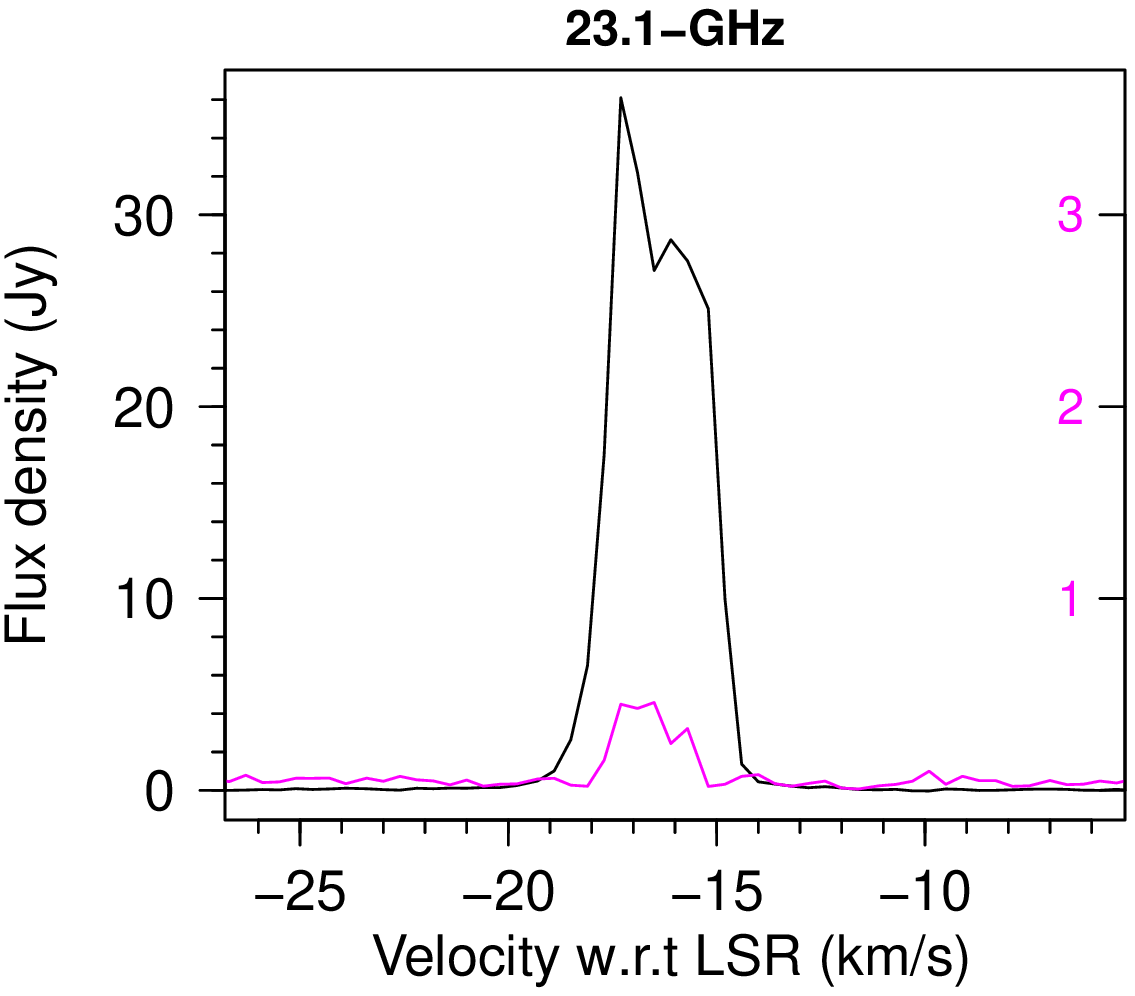,scale=0.51,angle=270}\hspace{-0.4cm}
\epsfig{figure=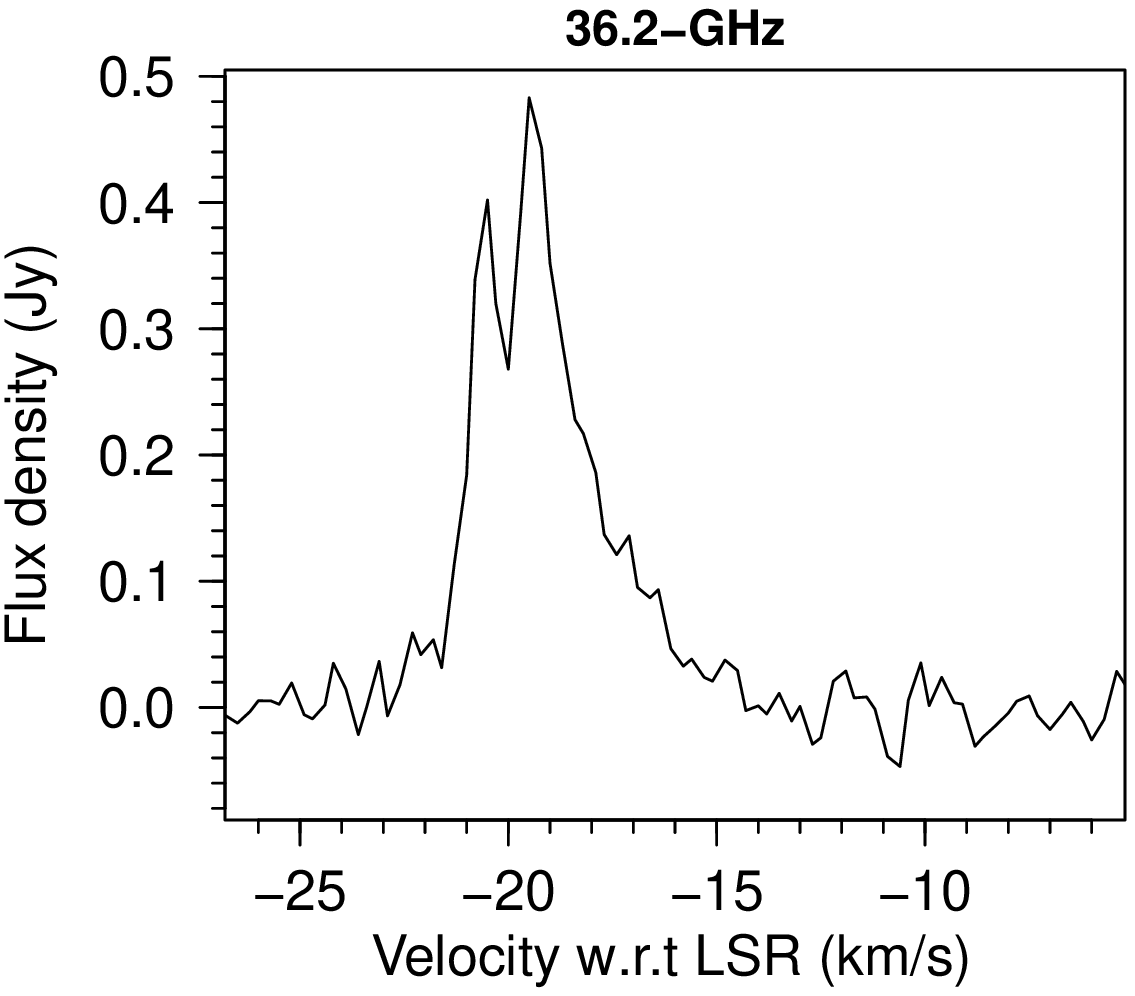,scale=0.51,angle=270}\hspace{-0.4cm}
\epsfig{figure=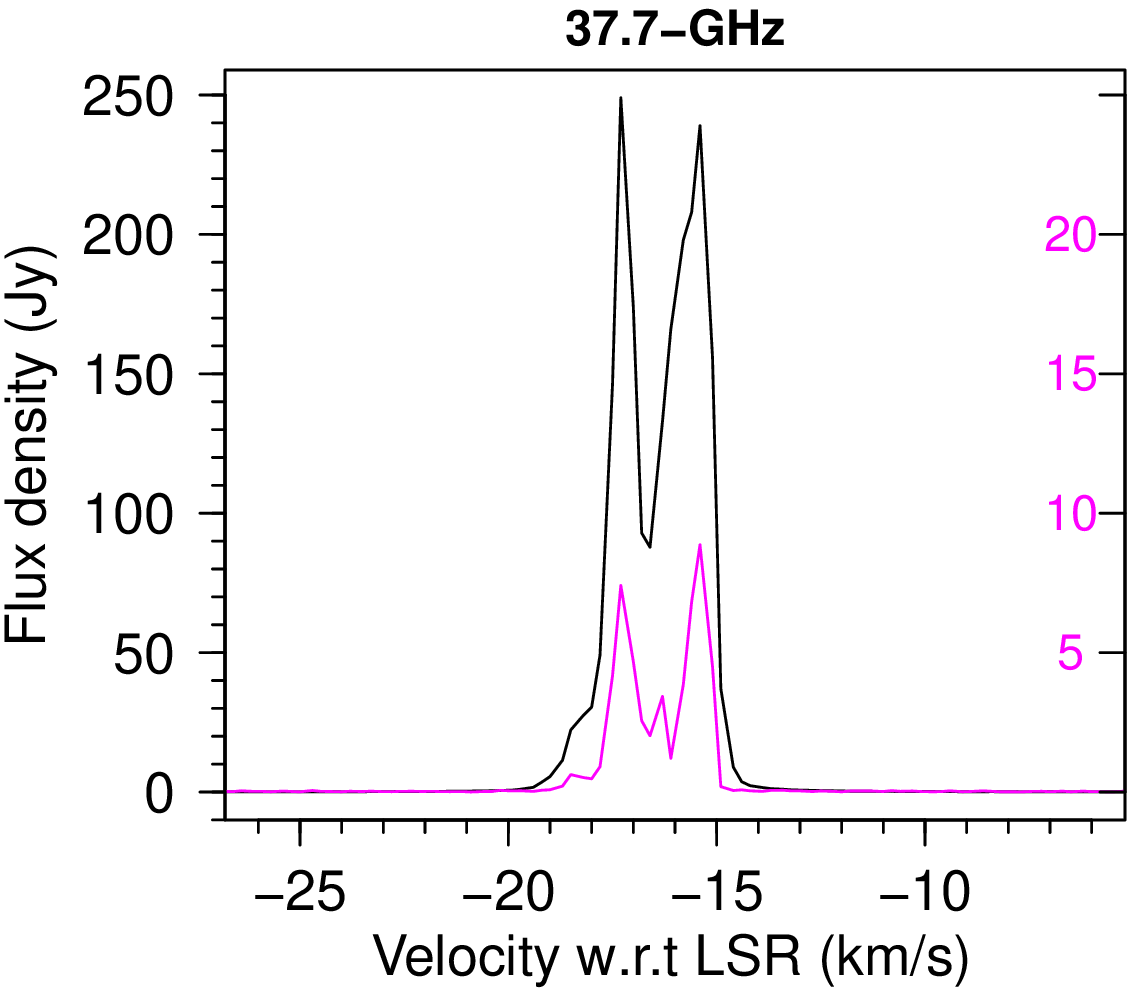,scale=0.51,angle=270}\vspace{-1cm}
\epsfig{figure=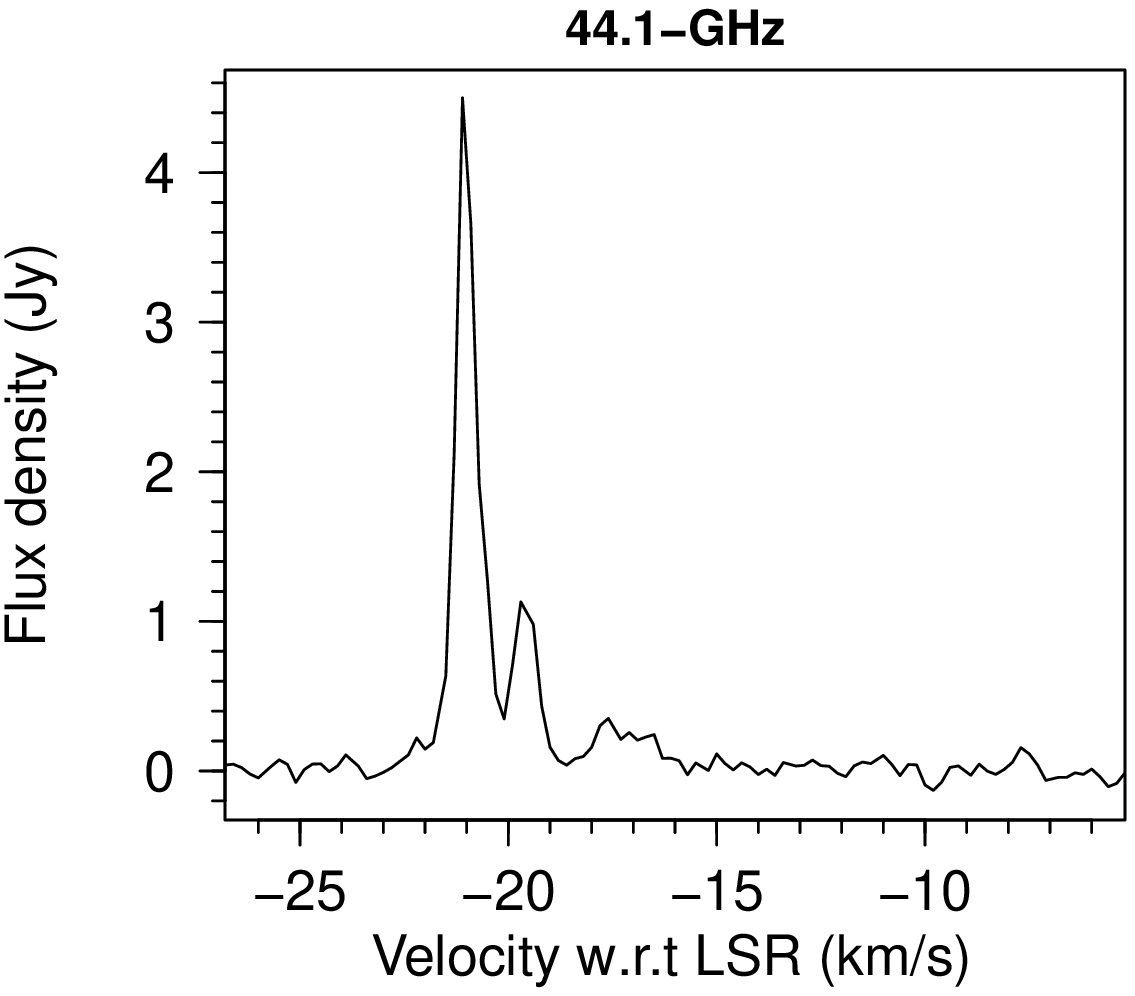,scale=0.51,angle=270}\hspace{-0.4cm}
\epsfig{figure=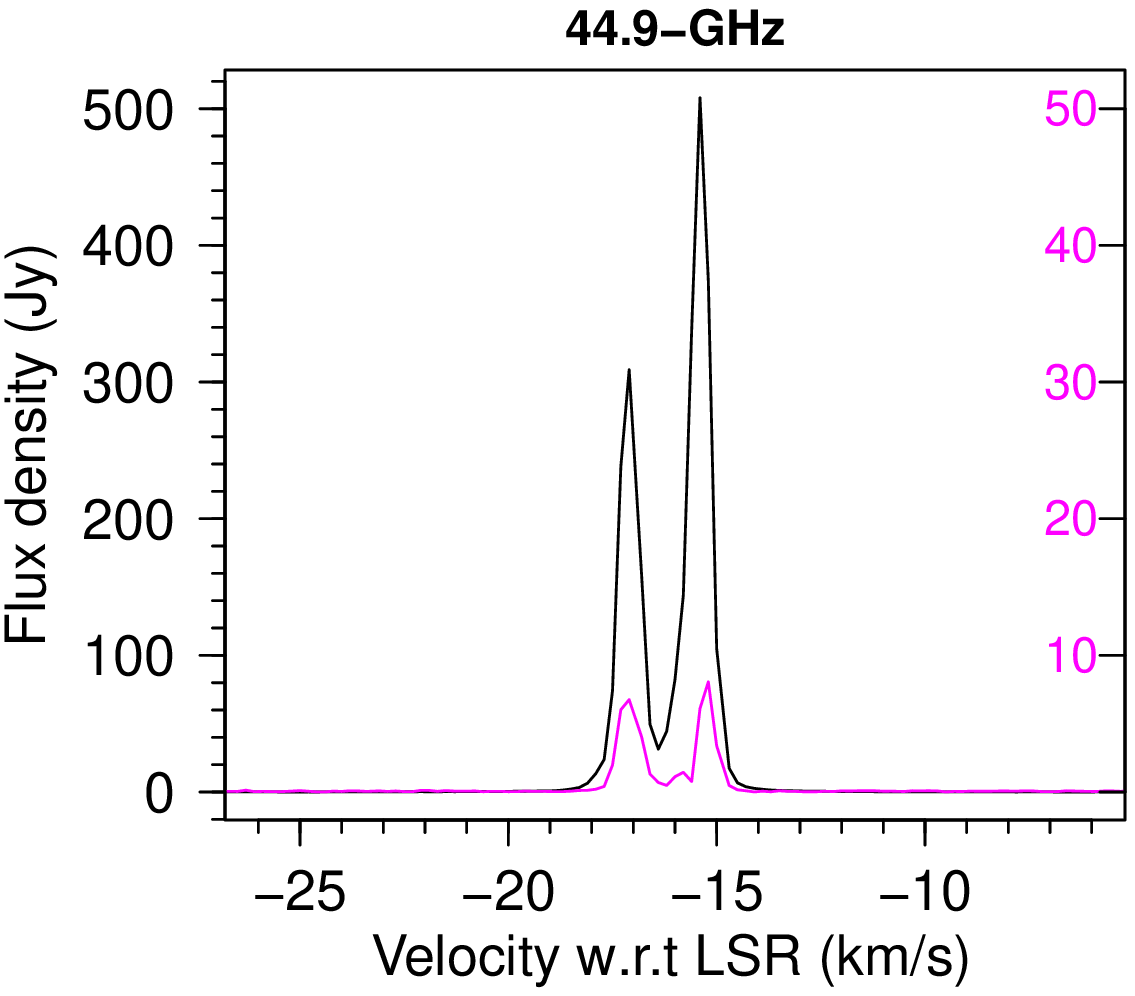,scale=0.51,angle=270}\hspace{-0.4cm}
\epsfig{figure=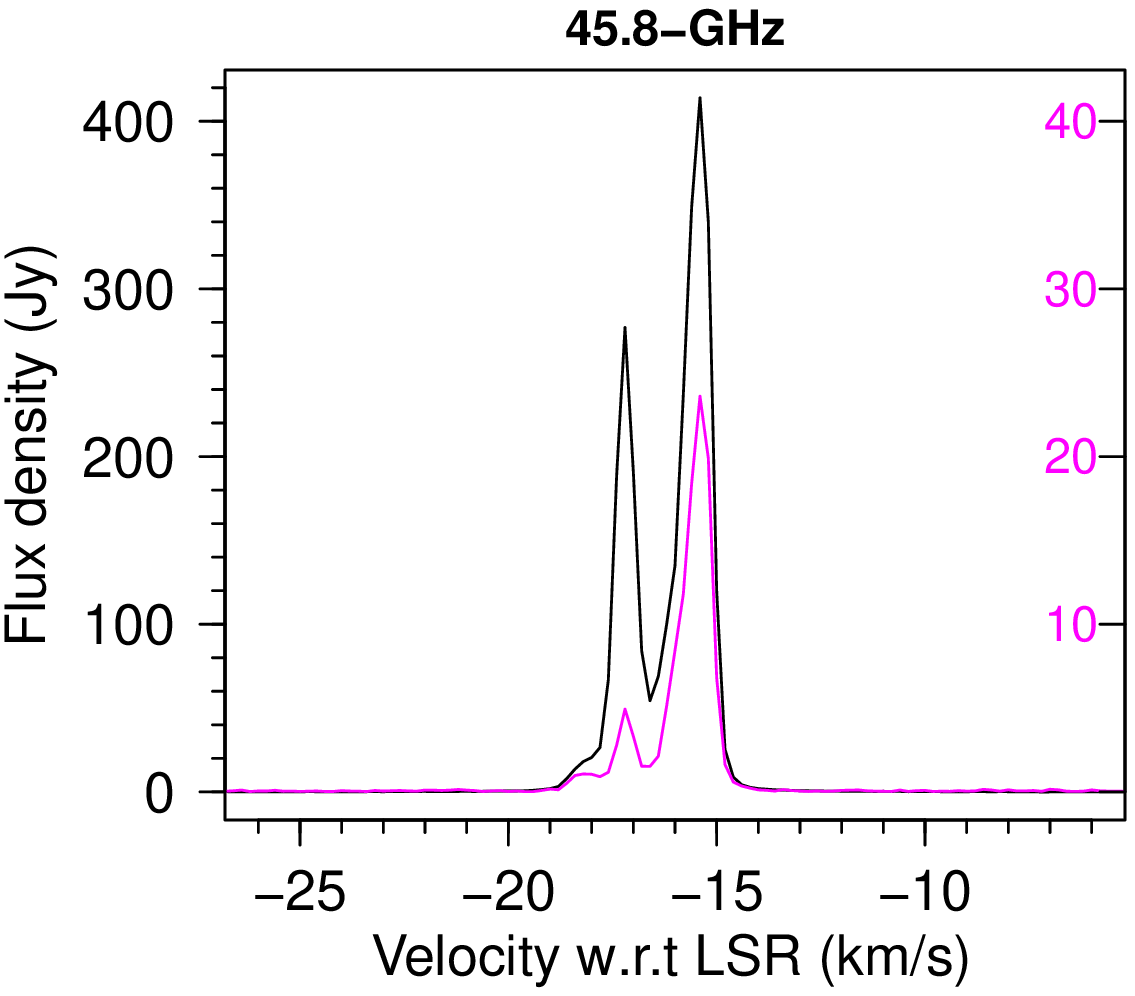,scale=0.51,angle=270}

\caption{Stokes I (black) and linearly polarised intensity ($\times$10 -- scale on right of plot; magenta) spectra of the methanol maser lines detected towards G\,358.931$-$0.030. Note that the 19.9-, 36.2- and 44.1-GHz transitions have no detected linearly polarised emission.}
\label{fig:spect}
\end{figure*}

Spectra of each of the detected lines are shown in Fig.~\ref{fig:spect} and include both Stokes I and the linearly polarized emission (scaled by a factor of 10). Despite being co-spatial, some variation in the velocity of the detected maser features can be seen across the transitions. These can all be accounted for by the uncertainty in the adopted rest frequencies, which are shown in Tables~\ref{tab:obs} and~\ref{tab:detections}. Comparison between the velocities show that the largest deviations from the 6.68-GHz maser velocity is in the 6.18-, 7.68- and 7.83-GHz lines. We suggest that refinements of the respective rest frequencies to 6181.146, 7682.246 and 7830.848~MHz would result in much closer velocity correspondence between all of the class II methanol maser transitions.

The derived positions listed in Table~\ref{tab:detections} confirm that all of the detected class II methanol maser lines are co-spatial with the 6.68-GHz methanol maser, and show a slight ($\sim$1\arcsec) variation from the published MMB position \citep{Caswell-2010}. The MMB position for the 6.68-GHz methanol maser was measured when the peak flux density was 6~Jy, a factor of $>$100 lower than for the current observations, which combined with the inferior receiver system available at the time of the observations can likely account for the discrepancy. Imaging of the 6.68-GHz transition being undertaken with other interferometers will provide further independent estimates of the absolute position in the near future. The largest deviation in our measured class II positions is in the weakest transition at 19.9-GHz, which is $\sim$0.5\arcsec\/ from our measured location of the 6.68-GHz maser. The location of the peak class I methanol maser emission in both the 36.2- and 44.1-GHz transitions are offset by $\sim$1.5\arcsec\/ from the class II maser position, and the precise correspondence between the 36.2- and 44.1-GHz lines indicate that this offset is genuine.

\begin{table*}
 \caption{Properties of the methanol masers detected towards G\,358.931$-$0.030, including the measured positions, the minimum, maximum and peak velocity, uncertainty in velocity (based on the uncertainty in the respective rest frequencies), the peak and integrated flux densities, as well as the maximum linear  and circular polarisation percentage. The newly discovered class~II maser transitions are indicated in bold font.}
  \begin{tabular}{lllllllcccl} \hline
 \multicolumn{1}{l}{\bf Spectral line} & {\bf RA (2000)} & {\bf Dec. (2000)}& {\bf V$_{min}$} & {\bf V$_{max}$} & {\bf V$_{peak}$} & {\bf V$_{Uncert.}$}& {\bf S$_{peak}$} & {\bf S$_{int}$} & {\bf linear } & {\bf circular }\\ 
  &\multicolumn{1}{c}{\bf ($^{h}$ $^{m}$ $^s$)} & \multicolumn{1}{c}{\bf ($^{\circ}$ $'$ $''$)}&\multicolumn{4}{c}{(\kms)}& {\bf (Jy)} & \bf{(Jy~\kms)} & {\bf pol. \%} & {\bf pol. \%}\\ \hline

{\bf 6.18-GHz}  &     17 43 10.10     &   $-$29 51 45.6  & $-$18.8 & $-$15.5 & $-$16.2 & (1.02) & 290 & 340 & 7.0 & $<$0.5\\ 
6.68-GHz &      17 43 10.10     &   $-$29 51 45.5  & $-$19.6 & $-$12.7 & $-$17.2 & (0.04) & 981 & 1110 & 7.5 & 0.5\\ 
{\bf 7.68-GHz} &      17 43 10.10     &   $-$29 51 45.6  & $-$20.3 & $-$13.4 & $-$15.9 & (1.95) &567 & 722 & 3.5 & $<$0.5\\ 
{\bf 7.83-GHz}  &     17 43 10.10     &   $-$29 51 45.6  & $-$19.3 & $-$12.6 & $-$16.6 & (1.92) & 571 & 711 & 3.5 & $<$0.5\\ 
\\

19.9-GHz &      17 43 10.14     &   $-$29 51 45.9  & $-$18.1 & $-$14.8 & $-$17.6 & (0.003) &0.3 & 0.7 & $<$0.5 & $<$0.5\\ 
{\bf 20.9-GHz}   &    17 43 10.12     &   $-$29 51 45.8  & $-$33.5 & $-$0.7 & $-$15.6 & (0.71) &978 & 1896 & 7.0 & $<$0.5 \\ 
23.1-GHz &      17 43 10.08     &   $-$29 51 45.6  & $-$19.8 & $-$13.2 & $-$17.3 & (0.006) &36.1 & 89.1 & 1.5 & $<$0.5\\ 
\\
36.2-GHz  &   17 43 09.89  & $-$29 51 45.8 & $-$21.3 & $-$17.1 & $-$19.5 & (0.12) &0.5 & 1.2 & $<$0.5 & $<$0.5\\
37.7-GHz	& 17 43 10.10 & $-$29 51 45.3 & $-$22.3 & $-$10.8 & $-$17.3 & (0.10) &249 & 496 & 3.5 & 0.5\\ 
\\
44.1-GHz 	    &	17 43 09.92 & $-$29 51 46.0	& $-$22.4 & $-$16.5 & $-$21.1 & (0.07) & 4.5 & 4.4 &$<$0.5 & $<$0.5\\ 

{\bf 44.9-GHz}  &         17 43 10.11 & $-$29 51 45.4 & $-$22.1 & $-$10.1 & $-$15.4 & (0.33) & 508 & 536 & 2.5 & $<$0.5\\ 
{\bf 45.8-GHz}  &         17 43 10.10 & $-$29 51 45.4 & $-$22.4 & $-$9.8 & $-$15.4 & (0.33) & 414 & 555 & 7.0 & 1.5\\ 
\hline

\end{tabular}\label{tab:detections}
\end{table*}


\section{Discussion}

\subsection{Newly discovered v$_t$=1 methanol masers at 6.18-, 20.9- and 44.9-GHz}

The clear detection of torsionally excited methanol maser lines resolves a significant outstanding uncertainty with current maser models, unambiguously demonstrating that pumping through the torsionally excited levels occurs, as required by the models for the strong, commonly observed transitions. We detected three torsionally excited methanol masers with rest frequencies near 6.18-, 20.9- and 44.9-GHz; the latter two were included in maser models and have been listed as maser candidates under some model conditions \citep[e.g.][]{Cragg-2005}. However, our current data shows much higher intensities than the models predict and therefore the current observations offer substantial new information for methanol maser pumping models.

Previous observations of the 20.9- and 44.9-GHz transitions have detected a handful of emission sites: \citet{Menten-1986} targeted 9 sources for the 20.9-GHz transition, and detected emission towards W3(OH), Orion KL and W51E, while \citet{Voronkov-2002} observed six strong class II methanol maser sites and Orion KL at 44.9-GHz, reporting detections toward Orion KL, W3(OH) and NGC\,6334F, and in both cases, W3(OH) showed some evidence of possible weak maser emission, although has never been confirmed. A search for the 44.9-GHz transition towards a sample of 272 sources with excess 4.5-$\mu$m emission in the {\em Spitzer} GLIMPSE survey has been conducted with the Mopra telescope and even though more than 40~\% of these sources have nearby 6.68-GHz methanol masers, no  
emission was detected to a limit of $\sim$2 Jy (Ellingsen et al., in prep.), demonstrating that strong emission in this transition is very rare.

The 6.18-GHz methanol transition was not included in the published lists of class II methanol maser candidates \citep[see e.g.][for the most comprehensive list]{Cragg-2005}, and the only previously reported search toward the rich class II methanol maser site, G\,345.01+1.79, found no emission above a 3-$\sigma$ detection limit of 0.15~Jy \citep{Chipman-2016}. The detection of such a highly excited transition suggests that we are observing a previously unidentified pumping regime for the class~II methanol masers. Modelling of this line requires further development of the methanol molecule scheme of levels and transitions, beyond the data available at present.

\subsection{Newly discovered class II v$_t$=0 methanol maser transitions at 7.68-, 7.83- and 45.8-GHz}

All three of the newly discovered v$_t$=0 class II methanol masers (at 7.68-, 7.83- and 45.8-GHz) were predicted to exhibit maser emission under some conditions \citep{Cragg-2005}, but no maser emission has been reported in any sources. The only published search for any of these transitions was by \citet{Chipman-2016} who targeted the four of the most prominent sites of class II methanol maser emission (G\,339.88$-$1.26, G\,345.01+1.79, NGC6334F, and W48) in the 7.68- and 7.83-GHz lines, finding no emission above a 3-$\sigma$ detection limit of 0.11~Jy.

\subsection{The known class II methanol maser transitions at 6.68-, 19.9-, 23.1-, 37.7-, 38.3- and 38.5-GHz }

Our observations detected a strong 37.7-GHz methanol maser at the location of G\,358.931$-$0.030 but find no accompanying emission in the 38.3- and 38.5-GHz transitions to 3-$\sigma$ detection limits of 66~mJy. Recent sensitive, high-resolution observations of the 37.7-, 38.3- and 38.5-GHz methanol maser lines towards 11 known 37.7-GHz sources found that all 37.7-GHz methanol masers stronger than 16~Jy had accompanying emission in both the 38.3- and 38.5-GHz transitions, with ratios ranging from 300:1 and 400:1 (in G\,339.884−1.259) to 0.31:1 and 0.26:1 (in NGC\,6334F) \citep{Ellingsen-2018}. The lower limit of the 37.3-GHz to 38.3- and 38.5-GHz flux density ratio in G\,358.931$-$0.030 is $\sim$377:1, indicating that the conditions in this source are at the extreme end of known sources exhibiting 37.7-GHz maser emission.

 The 6.68-GHz methanol maser was discovered in the MMB survey with a peak flux density of 10~Jy \citep{Caswell-2010} and was reported by \citet{Coconuts-2019} to be undergoing a period of bursting activity on 2019 Jan 21 (when it had a peak flux density of 19~Jy, quickly rising to 99 Jy by Jan 26). Our observations, 37 days later, now measure a peak flux density of 981~Jy. The report of flaring in the 6.68-GHz maser emission triggered followup observations of a number of different known class~II methanol maser transitions with the 19.9- and 23.1-GHz transitions having been detected by other telescopes prior to our observations by the M2O collaboration.  The current observations confirm that both of these transitions arise from the same region as the 6.68-GHz methanol masers.  

\subsection{Class I methanol masers at 36.2- and 44.1-GHz}

This is the first reported detection of class I methanol maser emission in the 36.2-GHz transition towards G\,358.931$-$0.030. We also report on our observations of the 44.1-GHz transition, which also independently detected by other members of the M2O collaboration (Kim et al. in prep). Although class I methanol masers in the 36.2- and 44.1-GHz lines are relatively common in the vicinity of young high-mass stars harboring 6.68-GHz masers \citep[e.g.][]{Voronkov-2014}, observations of these lines only cover a small fraction of the Galaxy. Without pre-burst observations we are unable to conclude whether or not these masers have appeared due to the ongoing flaring event or how they have been effected.

\subsection{Class II methanol maser polarisation}
We detect modest levels of linearly polarised emission in the 6.18-, 6.68-, 7.68-. 7.83-, 20.9-, 23.1-, 37.7-, 44.9- and 45.8-GHz transitions, with maximum polarised components in the range from 1.5 to 7.5\% (listed in Table~\ref{tab:detections}). We additionally detect low levels of circular polarisation in the 6.68-, 37.7- and 45.8-GHz transitions, in the 0.5 - 1.5~\% range. The levels of linear polarisation we detect are similar to those found in 6.68-GHz class II methanol masers by previous studies \citep[a few to $\sim$10\%, e.g.][]{Stack-2011,Surcis-2019}, who similarly find that circularly polarised components are scarce. Given the limitations of the current observations, further polarisation observations will be needed to determine a more thorough picture.

\subsection{Class II methanol maser flaring events}
\citet{Caswell-1995} observed more than 200 6.68-GHz methanol masers at multiple epochs over a period of approximately 18 months and concluded that while variability of class~II methanol masers was common, it was rarely large in amplitude, and, furthermore that higher intensity sources were generally less variable.  While a comparison of the spectra of a sample of strong 6.68-GHz methanol masers over a 20 year time baseline was used to estimate that individual spectral components have a typical lifetime of around 150 years \citep{Ellingsen-2007}.  These and other studies of the variability of class~II methanol masers resulted in the general consensus that the large amplitude, rapid variability that is relatively common in 22-GHz water masers \citep[e.g.][]{Felli-2007} is either absent, or rare in class~II methanol masers.  However, an increase in the active monitoring of 6.68-GHz methanol masers has detected several examples of extreme flaring in class~II methanol masers \citep[e.g.][]{Fugisawa-2015,MacLeod-2018}.  At the time of writing, the flare in the class~II maser emission toward G\,358.931$-$0.030 is ongoing and the timely sharing of results and collaboration within the maser community has resulted in an unprecedented wealth of data covering many different molecular transitions.  Previous class~II methanol flares have generally only been observed in one or both of the 6.68 and 12.2-GHz methanol transitions.  The flare in G\,358.931$-$0.030 has demonstrated that where extreme and rapid variability is seen in the 6.68-GHz transition, it may be accompanied by emission in other much rarer class~II methanol transitions.  We encourage further multi-wavelength studies of this source to complement the molecular studies and aid in their interpretation.

Flaring masers offer a particularly good opportunity to test maser pumping theories and detect new and uncommon masers due to the relatively rapid changes in the radiation field. The pumping cycles of methanol masers are significant and time is required to achieve equilibrium in the maser level populations \citep[e.g.][]{Sobolev-1994}. This means that in the case where the radiation field is changing rapidly, we are dealing with masers in the non-stationary regime where some new and previously unpredicted maser transitions, like the one at 6.18-GHz, can occur. These effects are not yet studied and their significance will be elucidated by theoretical modelling based on results of an ongoing monitoring effort of the detections reported here. 

The refinement of the maser pumping models will allow us to use sensitive observations covering multiple maser transitions to infer physical conditions in the masing gas. In particular, the less common transitions are generally inverted over a narrower range of physical conditions in pumping models and where emission is shown to be co-spatial, the relative intensity of the different transitions will significantly constrain the possible physical conditions.  A number of such studies have been undertaken towards high-mass star formation regions which exhibit a large number of class~II methanol maser transitions \citep[e.g.][]{Cragg-2001,Sutton-2001,Cragg-2005}.  However, they combined data collected over a period of many years using multiple telescopes (generally single-dish instruments) and so interpretation was limited by uncertainty due to potential variability and a lack of information as to the degree to which the emission from the different transitions was co-spatial.  Our observations show that the 10 class~II methanol maser transitions we have detected are co-spatial to around 0.2\arcsec (which is within the systematic positional uncertainty of the observations).  The intense emission in such a large number of different class~II methanol maser transitions presents a unique opportunity to undertake milliarcsecond-scale observations (using VLBI) and determine the degree to which they are co-spatial on milliarcsecond scales.  Where that is found to be the case, detailed maser modelling can potentially infer the physical conditions at unprecedented resolution for high-mass star formation regions - scales of a few AU for individual maser spots and variations in the conditions on scales of tens to hundreds of AU (the size of the maser clusters).

\section{Summary}

We have discovered six new class II methanol maser transitions towards G\,358.931$-$0.030, a 6.68-GHz methanol maser that was recently reported to be undergoing a period of flaring. These discoveries include three transitions in the torsionally excited (v$_t$=1) state, the 6.18-GHz 17$_{−2}$ $\rightarrow$ 18$_{−3}$ E (v$_t$=1), 20.9-GHz 10$_1$ $\rightarrow$ 11$_2$ A$^+$ (v$_t$=1) and 44.9-GHz 2$_0$ $\rightarrow$ 3$_1$ E (v$_t$=1) transitions, and comprise the first torsionally maser lines ever definitively detected. The detection of these lines validates predictions of methanol maser pumping models, but have unexpectedly high flux densities and so provide significant new information to refine current methanol maser pumping theories. 

In addition to the torsionally excited lines, we have discovered three new class II methanol masers in the 7.86-GHz 12$_4$ $\rightarrow$ 13$_3$ A$^-$ (v$_t$=0), 7.83-GHz 12$_4$ $\rightarrow$ 13$_3$ A$^+$ (v$_t$=0) and 5.8-GHz 9$_3$ $\rightarrow$ 10$_2$ E (v$_t$=0) transitions, and the first detections of the 4$_{-1}$ $\rightarrow$ 3$_0$ E  36.2-GHz and 7$_{-2}$ $\rightarrow$ 8$_{-1}$ E (v$_t$=0) 37.7-GHz. We present upper limits on the 6$_2$ $\rightarrow$ 5$_3$ A$^-$ and A$^+$ transitions at 38.3- and 38.5-GHz and show that the detected 37.7-GHz maser has an usually high flux density to exhibit no accompanying emission in the 38.3- and 38.5-GHz transitions. 

We suggest refinements of the 6.18-, 7.68- and 7.83-GHz rest frequencies to 6181.146, 7682.246 and 7830.848~MHz, respectively in order to achieve velocity correspondence with the 6.68-GHz methanol maser emission. 

The relative total, linearly and circularly polarised intensities of the large number of detected maser lines will be used in combination with the plethora of multiwavelenth observations (including infrared, maser VLBI and maser monitoring observations) to both inform maser pumping schemes as well as to infer the usual physical conditions associated with the flaring event.

\section*{Acknowledgments}

The Australia Telescope Compact Array and Mopra radio telescope are part of the Australia Telescope National Facility. This research has made use of NASA's Astrophysics
Data System Abstract Service. S.P.E. acknowledges the support of ARC Discovery Project (project number DP180101061). AMS was supported by the Russian Science Foundation (grant 18-12-00193).


\begin{thebibliography}{}
\expandafter\ifx\csname natexlab\endcsname\relax\def\natexlab#1{#1}\fi
\providecommand{\url}[1]{\href{#1}{#1}}

\bibitem[{{Batrla} {et~al.}(1987){Batrla}, {Matthews}, {Menten}, \&
  {Walmsley}}]{Batrla-1987}
{Batrla}, W., {Matthews}, H.~E., {Menten}, K.~M., \& {Walmsley}, C.~M. 1987,
  \nat, 326, 49

\bibitem[{{Breen} {et~al.}(2019){Breen}, {Contreras}, {Dawson}, {Ellingsen},
  {Voronkov}, \& {McCarthy}}]{Breen-2019}
{Breen}, S.~L., {Contreras}, Y., {Dawson}, J.~R., {et~al.} 2019, \mnras, 484,
  5072

\bibitem[{{Breen} {et~al.}(2016){Breen}, {Ellingsen}, {Caswell}, {Green},
  {Voronkov}, {Avison}, {Fuller}, \& {Quinn}}]{Breen-2016}
{Breen}, S.~L., {Ellingsen}, S.~P., {Caswell}, J.~L., {et~al.} 2016, \mnras,
  459, 4066

\bibitem[{{Breen} {et~al.}(2012){Breen}, {Ellingsen}, {Caswell}, {Green},
  {Voronkov}, {Fuller}, {Quinn}, \& {Avison}}]{Breen-2012a}
---. 2012, \mnras, 421, 1703

\bibitem[{{Breen} {et~al.}(2013){Breen}, {Ellingsen}, {Contreras}, {Green},
  {Caswell}, {Stevens}, {Dawson}, \& {Voronkov}}]{Breen-2013}
{Breen}, S.~L., {Ellingsen}, S.~P., {Contreras}, Y., {et~al.} 2013, \mnras,
  435, 524

\bibitem[{{Breen} {et~al.}(2015){Breen}, {Fuller}, {Caswell}, {Green},
  {Avison}, {Ellingsen}, {Gray}, {Pestalozzi}, {Quinn}, {Richards}, {Thompson},
  \& {Voronkov}}]{Breen-2015}
{Breen}, S.~L., {Fuller}, G.~A., {Caswell}, J.~L., {et~al.} 2015, \mnras, 450,
  4109
  
\bibitem[{{Burns} {et~al.}(2019){Burns}}]{Burns-2019}
{Burns}, R.~A., {Orosz}, G., {Bayandina}, O., {et~al.} 2019, \mnras, submitted
  
  

\bibitem[{{Caswell}(1997)}]{Caswell-1997}
{Caswell}, J.~L. 1997, \mnras, 289, 203

\bibitem[{{Caswell} {et~al.}(1995){Caswell}, {Vaile}, {Ellingsen}, {Whiteoak},
  \& {Norris}}]{Caswell-1995}
{Caswell}, J.~L., {Vaile}, R.~A., {Ellingsen}, S.~P., {Whiteoak}, J.~B., \&
  {Norris}, R.~P. 1995, \mnras, 272, 96

\bibitem[{{Caswell} {et~al.}(2000){Caswell}, {Yi}, {Booth}, \&
  {Cragg}}]{Caswell-2000}
{Caswell}, J.~L., {Yi}, J., {Booth}, R.~S., \& {Cragg}, D.~M. 2000, \mnras,
  313, 599

\bibitem[{{Caswell} {et~al.}(2010){Caswell}, {Fuller}, {Green}, {Avison},
  {Breen}, {Brooks}, {Burton}, {Chrysostomou}, {Cox}, {Diamond}, {Ellingsen},
  {Gray}, {Hoare}, {Masheder}, {McClure-Griffiths}, {Pestalozzi}, {Phillips},
  {Quinn}, {Thompson}, {Voronkov}, {Walsh}, {Ward-Thompson}, {Wong-McSweeney},
  {Yates}, \& {Cohen}}]{Caswell-2010}
{Caswell}, J.~L., {Fuller}, G.~A., {Green}, J.~A., {et~al.} 2010, \mnras, 404,
  1029

\bibitem[{{Chen} {et~al.}(2011){Chen}, {Ellingsen}, {Shen}, {Titmarsh}, \&
  {Gan}}]{Chen-2011}
{Chen}, X., {Ellingsen}, S.~P., {Shen}, Z.-Q., {Titmarsh}, A., \& {Gan}, C.-G.
  2011, \apjs, 196, 9

\bibitem[{{Chipman} {et~al.}(2016){Chipman}, {Ellingsen}, {Sobolev}, \&
  {Cragg}}]{Chipman-2016}
{Chipman}, A., {Ellingsen}, S.~P., {Sobolev}, A.~M., \& {Cragg}, D.~M. 2016,
  \pasa, 33, e056

\bibitem[{{Cragg} {et~al.}(2004){Cragg}, {Sobolev}, {Caswell}, {Ellingsen}, \&
  {Godfrey}}]{Cragg-2004}
{Cragg}, D.~M., {Sobolev}, A.~M., {Caswell}, J.~L., {Ellingsen}, S.~P., \&
  {Godfrey}, P.~D. 2004, \mnras, 351, 1327

\bibitem[{{Cragg} {et~al.}(2001){Cragg}, {Sobolev}, {Ellingsen}, {Caswell},
  {Godfrey}, {Salii}, \& {Dodson}}]{Cragg-2001}
{Cragg}, D.~M., {Sobolev}, A.~M., {Ellingsen}, S.~P., {et~al.} 2001, \mnras,
  323, 939

\bibitem[{{Cragg} {et~al.}(2005){Cragg}, {Sobolev}, \& {Godfrey}}]{Cragg-2005}
{Cragg}, D.~M., {Sobolev}, A.~M., \& {Godfrey}, P.~D. 2005, \mnras, 360, 533

\bibitem[{{Cyganowski} {et~al.}(2009){Cyganowski}, {Brogan}, {Hunter}, \&
  {Churchwell}}]{Cyganowski-2009}
{Cyganowski}, C.~J., {Brogan}, C.~L., {Hunter}, T.~R., \& {Churchwell}, E.
  2009, \apj, 702, 1615

\bibitem[{{Ellingsen}(2005)}]{Ellingsen-2005}
{Ellingsen}, S.~P. 2005, \mnras, 359, 1498

\bibitem[{{Ellingsen}(2007)}]{Ellingsen-2007}
---. 2007, \mnras, 377, 571

\bibitem[{{Ellingsen} {et~al.}(2011){Ellingsen}, {Breen}, {Sobolev},
  {Voronkov}, {Caswell}, \& {Lo}}]{Ellingsen-2011}
{Ellingsen}, S.~P., {Breen}, S.~L., {Sobolev}, A.~M., {et~al.} 2011, \apj, 742,
  109

\bibitem[{{Ellingsen} {et~al.}(2004){Ellingsen}, {Cragg}, {Lovell}, {Sobolev},
  {Ramsdale}, \& {Godfrey}}]{Ellingsen-2004}
{Ellingsen}, S.~P., {Cragg}, D.~M., {Lovell}, J.~E.~J., {et~al.} 2004, \mnras,
  354, 401

\bibitem[{{Ellingsen} {et~al.}(2003){Ellingsen}, {Cragg}, {Minier}, {Muller},
  \& {Godfrey}}]{Ellingsen-2003}
{Ellingsen}, S.~P., {Cragg}, D.~M., {Minier}, V., {Muller}, E., \& {Godfrey},
  P.~D. 2003, \mnras, 344, 73

\bibitem[{{Ellingsen} {et~al.}(2012){Ellingsen}, {Sobolev}, {Cragg}, \&
  {Godfrey}}]{Ellingsen-2012}
{Ellingsen}, S.~P., {Sobolev}, A.~M., {Cragg}, D.~M., \& {Godfrey}, P.~D. 2012,
  \apjl, 759, L5

\bibitem[{{Ellingsen} {et~al.}(2018){Ellingsen}, {Voronkov}, {Breen},
  {Caswell}, \& {Sobolev}}]{Ellingsen-2018}
{Ellingsen}, S.~P., {Voronkov}, M.~A., {Breen}, S.~L., {Caswell}, J.~L., \&
  {Sobolev}, A.~M. 2018, \mnras, 480, 4851

\bibitem[{{Felli} {et~al.}(2007){Felli}, {Brand}, {Cesaroni}, {Codella},
  {Comoretto}, {Di Franco}, {Massi}, {Moscadelli}, {Nesti}, {Olmi}, {Palagi},
  {Panella}, \& {Valdettaro}}]{Felli-2007}
{Felli}, M., {Brand}, J., {Cesaroni}, R., {et~al.} 2007, \aap, 476, 373

\bibitem[{{Fujisawa} {et~al.}(2015){Fujisawa}, {Yonekura}, {Sugiyama},
  {Horiuchi}, {Hayashi}, {Hachisuka}, {Matsumoto}, \&
  {Niinuma}}]{Fugisawa-2015}
{Fujisawa}, K., {Yonekura}, Y., {Sugiyama}, K., {et~al.} 2015, The Astronomer's
  Telegram, 8286

\bibitem[{{Green} {et~al.}(2009){Green}, {Caswell}, {Fuller}, {Avison},
  {Breen}, {Brooks}, {Burton}, {Chrysostomou}, {Cox}, {Diamond}, {Ellingsen},
  {Gray}, {Hoare}, {Masheder}, {McClure-Griffiths}, {Pestalozzi}, {Phillips},
  {Quinn}, {Thompson}, {Voronkov}, {Walsh}, {Ward-Thompson}, {Wong-McSweeney},
  {Yates}, \& {Cohen}}]{Green-2009}
{Green}, J.~A., {Caswell}, J.~L., {Fuller}, G.~A., {et~al.} 2009, \mnras, 392,
  783

\bibitem[{{Hunter} {et~al.}(2017){Hunter}, {Brogan}, {MacLeod}, {Cyganowski},
  {Chandler}, {Chibueze}, {Friesen}, {Indebetouw}, {Thesner}, \&
  {Young}}]{Hunter-2017}
{Hunter}, T.~R., {Brogan}, C.~L., {MacLeod}, G., {et~al.} 2017, \apjl, 837, L29

\bibitem[{{Hunter} {et~al.}(2018){Hunter}, {Brogan}, {MacLeod}, {Cyganowski},
  {Chibueze}, {Friesen}, {Hirota}, {Smits}, {Chandler}, \&
  {Indebetouw}}]{Hunter-2018}
{Hunter}, T.~R., {Brogan}, C.~L., {MacLeod}, G.~C., {et~al.} 2018, \apj, 854,
  170

\bibitem[{{Jordan} {et~al.}(2017){Jordan}, {Walsh}, {Breen}, {Ellingsen},
  {Voronkov}, \& {Hyland}}]{Jordan-2017}
{Jordan}, C.~H., {Walsh}, A.~J., {Breen}, S.~L., {et~al.} 2017, \mnras, 471,
  3915

\bibitem[{{Jordan} {et~al.}(2015){Jordan}, {Walsh}, {Lowe}, {Voronkov},
  {Ellingsen}, {Breen}, {Purcell}, {Barnes}, {Burton}, {Cunningham}, {Hill},
  {Jackson}, {Longmore}, {Peretto}, \& {Urquhart}}]{Jordan-2015}
{Jordan}, C.~H., {Walsh}, A.~J., {Lowe}, V., {et~al.} 2015, \mnras, 448, 2344

\bibitem[{{Kurtz} {et~al.}(2004){Kurtz}, {Hofner}, \&
  {{\'A}lvarez}}]{Kurtz-2004}
{Kurtz}, S., {Hofner}, P., \& {{\'A}lvarez}, C.~V. 2004, \apjs, 155, 149

\bibitem[{{Leurini} {et~al.}(2016){Leurini}, {Menten}, \&
  {Walmsley}}]{Leurini-2016}
{Leurini}, S., {Menten}, K.~M., \& {Walmsley}, C.~M. 2016, \aap, 592, A31

\bibitem[{{MacLeod} {et~al.}(2018){MacLeod}, {Smits}, {Goedhart}, {Hunter},
  {Brogan}, {Chibueze}, {van den Heever}, {Thesner}, {Banda}, \&
  {Paulsen}}]{MacLeod-2018}
{MacLeod}, G.~C., {Smits}, D.~P., {Goedhart}, S., {et~al.} 2018, \mnras, 478,
  1077

\bibitem[{{Menten}(1991)}]{Menten-1991}
{Menten}, K.~M. 1991, \apjl, 380, L75

\bibitem[{{Menten} {et~al.}(1986){Menten}, {Walmsley}, {Henkel}, {Wilson},
  {Snyder}, {Hollis}, \& {Lovas}}]{Menten-1986}
{Menten}, K.~M., {Walmsley}, C.~M., {Henkel}, C., {et~al.} 1986, \aap, 169, 271

\bibitem[{{Minier} {et~al.}(2003){Minier}, {Ellingsen}, {Norris}, \&
  {Booth}}]{Minier-2003}
{Minier}, V., {Ellingsen}, S.~P., {Norris}, R.~P., \& {Booth}, R.~S. 2003,
  \aap, 403, 1095

\bibitem[{{Moscadelli} {et~al.}(2017){Moscadelli}, {Sanna}, {Goddi},
  {Walmsley}, {Cesaroni}, {Caratti o Garatti}, {Stecklum}, {Menten}, \&
  {Kraus}}]{Moscadelli-2017}
{Moscadelli}, L., {Sanna}, A., {Goddi}, C., {et~al.} 2017, \aap, 600, L8

\bibitem[{{M{\"u}ller} {et~al.}(2004){M{\"u}ller}, {Menten}, \&
  {M{\"a}der}}]{Muller-2004}
{M{\"u}ller}, H.~S.~P., {Menten}, K.~M., \& {M{\"a}der}, H. 2004, \aap, 428,
  1019

\bibitem[{{Pickett} {et~al.}(1998){Pickett}, {Poynter}, {Cohen}, {Delitsky},
  {Pearson}, \& {M{\"u}ller}}]{Pickett-1998}
{Pickett}, H.~M., {Poynter}, R.~L., {Cohen}, E.~A., {et~al.} 1998, \jqsrt, 60,
  883
  
\bibitem[{{Rajabi} {et~al.}(2019){Rajabi}, {Houde}, {Bartkiewicz}, {Olech},
  {Szymczak}, \& {Wolak}}]{Rajabi-2019}
{Rajabi}, F., {Houde}, M., {Bartkiewicz}, A., {et~al.} 2019, \mnras, 484, 1590


\bibitem[{{Rayner} {et~al.}(2000){Rayner}, {Norris}, \&
  {Sault}}]{atca_polarisation}
{Rayner}, D.~P., {Norris}, R.~P., \& {Sault}, R.~J. 2000, \mnras, 319, 484

\bibitem[{{Sault} {et~al.}(1995){Sault}, {Teuben}, \& {Wright}}]{miriad}
{Sault}, R.~J., {Teuben}, P.~J., \& {Wright}, M.~C.~H. 1995, in Astronomical
  Society of the Pacific Conference Series, Vol.~77, Astronomical Data Analysis
  Software and Systems IV, ed. R.~A. {Shaw}, H.~E. {Payne}, \& J.~J.~E.
  {Hayes}, 433

\bibitem[{{Sobolev} {et~al.}(1997{\natexlab{a}}){Sobolev}, {Cragg}, \&
  {Godfrey}}]{Sobolev-1997a}
{Sobolev}, A.~M., {Cragg}, D.~M., \& {Godfrey}, P.~D. 1997{\natexlab{a}}, \aap

\bibitem[{{Sobolev} {et~al.}(1997{\natexlab{b}}){Sobolev}, {Cragg}, \&
  {Godfrey}}]{Sobolev-1997b}
---. 1997{\natexlab{b}}, \mnras

\bibitem[{{Sobolev} \& {Deguchi}(1994)}]{Sobolev-1994}
{Sobolev}, A.~M., \& {Deguchi}, S. 1994, \aap, 291, 569

\bibitem[{{Sobolev} \& {Parfenov}(2018)}]{Sobolev-2018}
{Sobolev}, A.~M., \& {Parfenov}, S.~Y. 2018, in IAU Symposium, Vol. 336,
  Astrophysical Masers: Unlocking the Mysteries of the Universe, ed.
  A.~{Tarchi}, M.~J. {Reid}, \& P.~{Castangia}, 57--58

\bibitem[{{Stack} \& {Ellingsen}(2011)}]{Stack-2011}
{Stack}, P.~D., \& {Ellingsen}, S.~P. 2011, \pasa, 28, 338

\bibitem[{{Sugiyama} {et~al.}(2019){Sugiyama}, {Saito}, {Yonekura}, \&
  {Momose}}]{Coconuts-2019}
{Sugiyama}, K., {Saito}, Y., {Yonekura}, Y., \& {Momose}, M. 2019, The
  Astronomer's Telegram, 12446

\bibitem[{{Surcis} {et~al.}(2019){Surcis}, {Vlemmings}, {van Langevelde},
  {Hutawarakorn Kramer}, \& {Bartkiewicz}}]{Surcis-2019}
{Surcis}, G., {Vlemmings}, W.~H.~T., {van Langevelde}, H.~J., {Hutawarakorn
  Kramer}, B., \& {Bartkiewicz}, A. 2019, \aap, 623, A130

\bibitem[{{Sutton} {et~al.}(2001){Sutton}, {Sobolev}, {Ellingsen}, {Cragg},
  {Mehringer}, {Ostrovskii}, \& {Godfrey}}]{Sutton-2001}
{Sutton}, E.~C., {Sobolev}, A.~M., {Ellingsen}, S.~P., {et~al.} 2001, \apj,
  554, 173

\bibitem[{{Titmarsh} {et~al.}(2016){Titmarsh}, {Ellingsen}, {Breen}, {Caswell},
  \& {Voronkov}}]{Titmarsh-2016}
{Titmarsh}, A.~M., {Ellingsen}, S.~P., {Breen}, S.~L., {Caswell}, J.~L., \&
  {Voronkov}, M.~A. 2016, \mnras, 459, 157

\bibitem[{{Tsunekawa} {et~al.}(1995){Tsunekawa}, {Ukai}, {Toyama}, \&
  {Takagi}}]{Tsunekawa-1995}
{Tsunekawa}, S., {Ukai}, T., {Toyama}, A., \& {Takagi}, K. 1995, in Technical
  report, Report for the Grant-in-aid for Scientific Research on Priority Areas
  (Interstellar Matter, 1991-1994) of the Ministry of Education, Science and
  Culture, Japan. Toyama University


\bibitem[{{Volvach} {et~al.}(2019){Volvach}, {Volvach}, \&
  {Larionov}}]{Volvach-2019}
{Volvach}, A.~E., {Volvach}, L.~N., \& {Larionov}, M.~G. 2019, Astronomy Letters, submitted


\bibitem[{{Voronkov} {et~al.}(2002){Voronkov}, {Austin}, \&
  {Sobolev}}]{Voronkov-2002}
{Voronkov}, M.~A., {Austin}, M.~C., \& {Sobolev}, A.~M. 2002, \aap, 387, 310

\bibitem[{{Voronkov} {et~al.}(2014){Voronkov}, {Caswell}, {Ellingsen}, {Green},
  \& {Breen}}]{Voronkov-2014}
{Voronkov}, M.~A., {Caswell}, J.~L., {Ellingsen}, S.~P., {Green}, J.~A., \&
  {Breen}, S.~L. 2014, \mnras, 439, 2584

\bibitem[{{Wilson} {et~al.}(2011){Wilson}, {Ferris}, {Axtens}, {Brown},
  {Davis}, {Hampson}, {Leach}, {Roberts}, {Saunders}, {Koribalski}, {Caswell},
  {Lenc}, {Stevens}, {Voronkov}, {Wieringa}, {Brooks}, {Edwards}, {Ekers},
  {Emonts}, {Hindson}, {Johnston}, {Maddison}, {Mahony}, {Malu}, {Massardi},
  {Mao}, {McConnell}, {Norris}, {Schnitzeler}, {Subrahmanyan}, {Urquhart},
  {Thompson}, \& {Wark}}]{Wilson-2011}
{Wilson}, W.~E., {Ferris}, R.~H., {Axtens}, P., {et~al.} 2011, \mnras, 416, 832

\bibitem[{{Xu} \& {Lovas}(1997)}]{Xu-1997}
{Xu}, L.-H., \& {Lovas}, F.~J. 1997, Journal of Physical and Chemical Reference
  Data, 26, 17

\bibitem[{{Xu} {et~al.}(2008){Xu}, {Li}, {Hachisuka}, {Pandian}, {Menten}, \&
  {Henkel}}]{Xu-2008}
{Xu}, Y., {Li}, J.~J., {Hachisuka}, K., {et~al.} 2008, \aap, 485, 729

\bibitem[{{Yusef-Zadeh} {et~al.}(2013){Yusef-Zadeh}, {Cotton}, {Viti},
  {Wardle}, \& {Royster}}]{YZ-2013}
{Yusef-Zadeh}, F., {Cotton}, W., {Viti}, S., {Wardle}, M., \& {Royster}, M.
  2013, \apjl, 764, L19

\bibitem[{{Zinchenko} {et~al.}(2017){Zinchenko}, {Liu}, {Su}, \&
  {Sobolev}}]{Zinchenko-2017}
{Zinchenko}, I., {Liu}, S.-Y., {Su}, Y.-N., \& {Sobolev}, A.~M. 2017, \aap,
  606, L6

\end{thebibliography}

\end{document}